\documentclass[aps,preprint]{revtex4}
\usepackage{amsfonts}
\usepackage{amssymb,epsfig}
\usepackage{latexsym}
\usepackage{amsmath}
\usepackage{graphicx}
\usepackage{url}

\begin{document}

\title{Interacting Ricci Dark Energy with Logarithmic Correction}

\author{Antonio Pasqua$^{1}$ \footnote{Email:toto.pasqua@gmail.com},~~A. Khodam-Mohammadi$^{2,3}$ \footnote{Email:khodam@basu.ac.ir},~
~Mubasher Jamil$^{4,5}$
\footnote{Email:mjamil@camp.nust.edu.pk}~and~R. Myrzakulov$^{6}$
\footnote{Email:rmyrzakulov@csufresno.edu}}
 \affiliation{$^{1}$ Department of Physics,
University of Trieste, Via Valerio, 2 34127 Trieste, Italy \\$^{2}$
Physics Department, Faculty of Science, Bu-Ali Sina University,
Hamedan 65178, Iran\\$^{3}$ Department of Physics, Faculty of
Science, Shiraz University, Shiraz 71454, Iran\\$^{4}$ Center for
Advanced Mathematics and Physics (CAMP), National University of
Sciences and Technology (NUST), H-12, Islamabad, Pakistan\\$^{5}$
Eurasian International Center for Theoretical Physics, Eurasian
National University, Astana 010008, Kazakhstan \\$^{6}$ Department
of Physics, California State University, Fresno, CA 93740 USA}
\begin{abstract}
Motivated by the holographic principle, it has been suggested that
the dark energy density  may be inversely proportional to the area
 $A$ of the event horizon of the universe. However, such a model would have a causality problem. In this work, we consider the
 entropy-corrected version of the holographic dark energy model in the non-flat FRW universe and we propose to replace the future
 event horizon area with the inverse of the Ricci scalar curvature. We obtain the equation of state (EoS) parameter $\omega_{\Lambda}$,
 the deceleration parameter $q$ and $\Omega_D'$ in the presence of interaction between Dark Energy (DE) and Dark Matter (DM). Moreover,
 we reconstruct the potential and the dynamics of the tachyon, K-essence, dilaton and quintessence scalar field models according to the
 evolutionary behavior of the interacting entropy-corrected holographic dark energy model.\\
\end{abstract}
\maketitle

\section{Introduction}
It is popularly believed among astrophysicists and cosmologists that our universe is experiencing an accelerated expansion.
The evidence of the accelerated expansion of the universe is proved by numerous and complementary cosmological observations,
 like the Supernovae Ia (SNIa) \citep{perlmutter,astier}, the Cosmic Microwave Background (CMB) anisotropies,
 observed mainly by WMAP (Wilkinson Microwave Anisotropy Probe) \citep{bennett-09-2003,spergel-09-2003}, the Large Scale
 Structure (LSS) \citep{tegmark,abz1,abz2} and X-ray \citep{allen} experiments. \\
In the framework of standard Freidmann-Robertson-Walker (FRW)
cosmology, a missing energy component with negative pressure (called
Dark Energy (DE)) is the source of this expansion. Careful analysis
of cosmological observations, in particular of WMAP data
\citep{spergel-09-2003,bennett-09-2003,peiris} indicates that the
two-thirds of the total energy of the universe is been occupied by
the DE whereas DM occupies almost the rest (the baryonic matter representing only a few percent of the total).\\
The nature of DE is still unknown, and scientists have proposed many
candidates in order to describe it
(see \citep{Padmanabhan-07-2003,peebles,copeland-2006} and references therein for good reviews).\\
The time-independent cosmological constant, $\Lambda$, as vacuum
energy density, with equation of state (EoS) parameter $\omega=-1$
is the earliest, most famous and simplest theoretical candidate for DE.\\
Cosmologists know that the cosmological constant suffers from two well-known difficulties: the fine-tuning and the
 cosmic coincidence problems \citep{copeland-2006}. The former asks why the vacuum energy density is so small
 (an order of $10^{-123}$ smaller than what we observe) \citep{weinberg} and the latter says why vacuum energy and DM are
 nearly equal today (which is an incredible coincidence if there are no internal connections between them).\\
The alternative candidates for DE problem are the dynamical DE scenarios with no longer constant but time-varying EoS, $\omega$.
 According to some analysis on the SNe Ia observational data, it has been shown that the time-varying DE models give a better fit
  compared with a cosmological constant. There are two different categories for dynamical DE scenarios: (i) scalar fields,
  including quintessence \citep{wetterich,ratra}, K-essence \citep{chiba-07-2000,armendariz-11-2000,armendariz-05-2001},
  phantoms \citep{caldwell-10-2002,nojiri-06-2003,nojiri-07-2003}, tachyon \citep{sen-04-2002,Padmanabhan-06-2002,Padmanabhan-10-2002},
   dilaton \citep{gasperini,piazza,arkani}, quintom \citep{elizalde-08-2004,nojiri-03-2005,anisimov} and so forth, and (ii)
   interacting DE models, including Chaplygin gas models \citep{Kamenshchik,setare-11-2007,bento}, braneworld models \citep{deffayet,sahni},
    holographic DE (HDE) \citep{cohen,horava-08-2000,setare-01-2007,setare-10-2007,setare-11-2006,setare-05-2007,setare-01-05-2007,setare-09-2007}
    and agegraphic DE (ADE) models \citep{cai-12-2007,wei-02-2008}. For a good review about
    the problem of DE, including a survey of some theoretical models, see \citep{miao}.\\
An important advance in the studies of black hole theory and string theory is the suggestion of the so called holographic principle.
 According to the holographic principle, the number of degrees of freedom of a physical system should be finite and should scale with
  its bounding area rather than with its volume \citep{thooft} and it should be constrained by an infrared cut-off \citep{cohen}.
  The holographic DE (HDE), based on the holographic principle proposed by \citep{fischler}, is one of the most interesting DE candidates
   and it has been widely studied in literature \citep{huang-08-2004,hsu,wang-09-2005,guberina-05-2005,guberina-05-2006,gong-09-2004,jamil-01-2010,J2011,
    SJ2011,she2011,sheykhi-01-2011,sheykhi-03-01-2010,elizalde-05-2005,setare-01-05-2007,setare-02-2010,2011setare,2010setarejamil,2011karami,2010farooq,
    SKJ2010,2010IJSF,2009JSS,karami-03-2010,sheykhi-11-2009,setare-09-2006,setare-01-2007,setare-08-2008,setare-11-2006,setare-11-2007,zhangX-08-2005,
    zhangX-11-2006,zhangX-07-2007,enqvist-02-2005,shen,jamilfarooq2009}. HDE models have also been constrained and tested by various astronomical
      observations \citep{zhangX-08-2005,zhangX-07-2007,huang-08-2004,enqvist-02-2005,wangY,micheletti,zhangX-05-2009,feng-02-2005,kao,shen} as well as
      by the anthropic principle \citep{huang-03-2005}.\\
Applying the holographic principle to cosmology, the upper bound of
the entropy contained in the universe can be obtained
\citep{fischler}.
 Following this line, \citep{li-12-2004} suggested the following constraint on its energy density:
\begin{eqnarray}
    \rho_{\Lambda}\leq3c^2M_p^2L^{-2}, \label{1}
\end{eqnarray}
where $c$ is a numerical constant, $L$ denotes the IR cut-off
radius, $M_p = (8\pi G)^{-1/2}$ is the reduced Planck mass ($G$
represents the gravitational constant) and the equality sign holds
only when the holographic bound is saturated. Obviously, in the
derivation of HDE, the black hole entropy $S_{BH}$ plays an
important role. As it is well known, $S_{BH} = A/(4G)$, where
$A\approx L^2$ is the area of horizon and $G$ is the gravitational
constant. However, in literature, this entropy-area relation can be
modified to
\citep{banerjee-04-2008,banerjee-06-2008,banerjee-05-2009,Majhi,jamil-03-2010,wei-02-2010,easson,sad2010,jam2011}:
\begin{eqnarray}
    S_{BH} = \frac{A}{4G}+\tilde{\alpha} \log \left( \frac{A}{4G} \right) + \tilde{\beta},\label{2}
\end{eqnarray}
where $\tilde{\alpha}$ and $\tilde{\beta}$ are dimensionless constants of order of unity. These corrections can appear in the black hole
 entropy in Loop Quantum Gravity (LQG). They can also be due to thermal equilibrium fluctuation, quantum fluctuation, or mass and charge
 fluctuations. The quantum corrections provided to the entropy-area relationship leads to the currvature correction in the Einstein-Hilbert
  action and viceversa \citep{zhu,cai-08-2009,nojiri-2001}. Using the corrected entropy-area relation (\ref{2}), the energy density of the
   entropy-corrected HDE (ECHDE) can be written as \citep{wei-10-2009}:
\begin{eqnarray}
    \rho_{\Lambda}=3\alpha M_p^2L^{-2}+ \gamma_1 L^{-4} \log \left( M_p^2L^2 \right) + \gamma_2 L^{-4},\label{3}
\end{eqnarray}
where $\gamma_1$ and $\gamma_2$ are dimensionless constants of order unity. In the limiting case $\gamma_1= \gamma_2 = 0$, Eq. (\ref{3})
yields the well-known HDE density.\\
The first term on Eq. (\ref{3}) corresponds to the usual holographic energy density. The second and the third terms are due to entropy
corrections: since they can be comparable to the first term only when $L$ is very small, the corrections given  by them make sense only
at the early evolutionary stage of the universe. When the universe becomes large, Eq. (\ref{3}) reduce to that of the ordinary HDE.\\
Inspired by the HDE models, in this paper we propose to consider another possibility: the IR cut-off radius $L$ is given by the average
radius of Ricci scalar curvature, $R^{-1/2}$, so that we have the DE density $\rho_{\Lambda} \propto R$. We remember that the Ricci scalar
can be written as:
\begin{eqnarray}
    R=6\left( \dot{H}+2H^2+\frac{k}{a\left(t\right)^2}\right),\label{4}
\end{eqnarray}
where $H=\dot{a}\left(t\right)/a\left( t \right)$ is the Hubble parameter, $\dot{H}$ is the derivative of the Hubble parameter with respect
to the cosmic time $t$, $a\left(t\right)$ is a dimensionless scale factor and $k$ is the curvature parameter which can assume the
 values $-1,\, 0,\, +1$ which yield, respectively, an open, a flat or a closed FRW universe. \\
Substituting $L$ with $R^{-1/2}$, we can write the energy density of Ricci ECHDE (R-ECHDE) as:
\begin{eqnarray}
    \rho_{\Lambda}=3\alpha M_p^2R+\gamma_1R^2 \log\left(M_p^2/R\right)+\gamma_2R^2,\label{5}
\end{eqnarray}
where $\alpha$, $\gamma_1$ and $\gamma_2$ are three constants, $M_p=\left(8\pi G\right)^{-1/2}$ is the modified Planck mass ($G$ is
the gravitational constant). many authors applied the entropy correction terms in an interacting/non-interacting and flat/non-flat
universe with modified IR-cutoff (for example see \citep{2011khodam,khodam1,khodam2}).  \\
This paper is organized as follows. In Section 2, we describe the physical contest we are working in and we derive the EoS
 parameter $\omega_{\Lambda}$, the deceleration parameter $q$ and $\Omega_D'$ for our model in a non-flat universe. In Section 3,
 we establish a correspondence between our model and the tachyon, K-essence, dilaton and quintessence fields. In Section 4 we write
 the Conclusions of this paper.

\section{INTERACTING MODEL IN A NON-FLAT universe}
Observational evidence suggest that our universe is not perfectly flat but it has a small positive curvature, which implies a closed universe.
 The tendency of a closed universe is shown in cosmological (in particular CMB) experiments \citep{bennett-09-2003,spergel-09-2003,tegmark,seljak,
 spergel-06-2007,sievers,netterfield,benoit-01-03-2003,benoit-02-03-2003}. Moreover, the measurements of the cubic correction to
 the luminosity-distance relation of Supernova measurements reveal a closed universe \citep{caldwell-09-2004, wang-01-2005}. For the above
 reasons, we prefer to consider a non-flat universe.\\
Within the framework of the standard Friedmann-Robertson-Walker (FRW) cosmology, the line element for non-flat universe is given by:
\begin{eqnarray}
    &&ds^2=\nonumber\\
    &&-dt^2+a^2(t)\left(\frac{dr^2}{1-kr^2} +r^2 \left(d\theta ^2 + \sin^2 \theta d\varphi ^2\right) \right)\label{6}
\end{eqnarray}
The corresponding Friedmann equation takes the form:
\begin{eqnarray}
    H^2+\frac{k}{a^2}=\frac{1}{3M^2_p}\left( \rho _{\Lambda} + \rho _{m}\right),\label{7}
\end{eqnarray}
where $\rho _{\Lambda}$ and $\rho _{m}$ are the energy densities of DE and DM, respectively.\\
We also define the fractional energy densities for matter, curvature and DE, respectively, as:
\begin{eqnarray}
    \Omega _m &=& \frac{\rho_m}{\rho _{cr}} = \frac{\rho _m}{3M_p^2H^2},\label{8}\\
    \Omega _k &=& \frac{\rho_k}{\rho _{cr}} = \frac{k}{H^2a^2},\label{9}\\
    \Omega _{\Lambda} &=& \frac{\rho_{\Lambda}}{\rho _{cr}} = \frac{\rho _{\Lambda}}{3M_p^2H^2},\label{10}
\end{eqnarray}
where $\rho_{cr} = 3M^2_pH^2$ represents the critical density. The parameter $\Omega _k$ represents the contribution to the total density
from the spatial curvature. Recent observations support a closed universe with a small positive curvature $\Omega _k \cong 0.02$ \citep{spergel-06-2007}.\\
Using Eqs. (\ref{8}), (\ref{9}) and (\ref{10}) it is possible to write the Friedmann Eq. (\ref{7}) in the following form:
\begin{eqnarray}
\Omega _{m} + \Omega _{\Lambda} = 1 + \Omega _k.\label{11}
\end{eqnarray}
In order to preserve the Bianchi identity or local energy-momentum conservation law, i.e. $\nabla_{\mu}T^{\mu \nu}=0$, the total energy density
$\rho_{tot} = \rho_{\Lambda} + \rho_m$ must satisfy the following equation:
\begin{eqnarray}
    \dot{\rho}_{tot}+3H\left( 1+\omega \right)\rho_{tot}=0\label{12},
\end{eqnarray}
where $\omega = p_{tot}/\rho_{tot}$ is the total EoS.
By assuming an interaction between DE and DM, the two energy densities $\rho_{\Lambda}$ and $\rho_m$ are conserved separately and the conservation
equations take the following form:
\begin{eqnarray}
    \dot{\rho}_{\Lambda}&+&3H\rho_{\Lambda}\left(1+w_{\Lambda}\right)=-Q, \label{13} \\
\dot{\rho}_m&+&3H\rho_m=Q .\label{14}
\end{eqnarray}
$Q$ represents the interaction term which can be, in general, an arbitrary function of cosmological parameters, like the Hubble parameter $H$
and energy densities $\rho_m$ and $\rho_{\Lambda}$, i.e. $Q(H\rho_m,H\rho_{\Lambda})$. The simplest choice for $Q$ is:
\begin{eqnarray}
    Q = 3b^2H(\rho _m + \rho _{\Lambda}),\label{15}
\end{eqnarray}
with $b^2$ a coupling parameter between DM and DE
\citep{amendola-08-2001,amendola-08-2002,setare-06-2010,sheykhi-01-2011,farooq,jamil-01-2010,j2010,
zimdahl-11-2001,zimdahl-03-2003,setare-02-2010} although more
general interaction terms can be used \citep{jamil-08-2008}.
However, since the nature of DM and DE remains unknown, different
Lagrangians have been proposed to generate this interaction term.
Positive values of $b^2$ indicate transition from DE to matter and
vice versa for negative values of $b^2$. Sometimes $b^2$ is taken in
the range [0,1] \citep{zhang-02-2006}. The case with $b^2 = 0$
represents the non-interacting FRW model while $b^2 = 1$ yields the
complete transfer of energy from DE to matter. Recently, it is
reported that this interaction is observed in the Abell cluster A586
showing a transition of DE into DM and vice versa
\citep{bertolami-10-2007,jamil-11-2008}.
However the strength of this interaction is not clearly identified \citep{feng-09-2007}.  \\
Observations of the CMB and of galactic clusters show that the coupling parameter $b^2 < 0.025$, i.e. a small but positive constant of order of the
unity \citep{ichiki,amendola-04-2007}. A negative coupling parameter is avoided due to violation of thermodynamical laws.\\
We now want to derive the expression for the EoS parameter
$\omega_{\Lambda}$ for our model.\\ Using (\ref{7}), the Ricci
scalar can be written as:
\begin{eqnarray}
    R=6\left(  \dot{H} + H^2 + \frac{\rho_m+\rho_{\Lambda}}{3M_p^2} \right). \label{17}
\end{eqnarray}
From the Friedmann Eq. (\ref{7}) we obtain that:
\begin{eqnarray}
    \dot{H}=\frac{k}{a^2}-\frac{1}{2M_p^2}\left(  \rho_m+\rho_{\Lambda}\left( 1 + \omega_{\Lambda} \right) \right). \label{18}
\end{eqnarray}
Adding Eqs. (\ref{7}) and (\ref{18}), we obtain:
\begin{eqnarray}
    \dot{H}+H^2=\frac{\rho_m+\rho_{\Lambda}}{3M_p^2}-\frac{1}{2M_p^2}\left(  \rho_m+\rho_{\Lambda}\left( 1 + \omega_{\Lambda} \right) \right). \label{19}
\end{eqnarray}
Therefore, the Ricci scalar can be written as:
\begin{eqnarray}   R=\frac{\rho_m+\rho_{\Lambda}}{M_p^2}-\frac{3\rho_{\Lambda}\omega_{\Lambda}}{M_p^2}.\label{20}
\end{eqnarray}
From Eq. (\ref{20}) we can easily derive the expression of the EoS parameter $\omega_{\Lambda}$:
\begin{eqnarray}    \omega_{\Lambda}=-\frac{RM_p^2}{3\rho_{\Lambda}}+\frac{\rho_{\Lambda}+\rho_{m}}{3\rho_{\Lambda}}=-\frac{RM_p^2}{3\rho_{\Lambda}}+\frac{\Omega_{\Lambda}+\Omega_{m}}{3\Omega_{\Lambda}}.\label{21}
\end{eqnarray}
Substituting the expression of the energy density given in Eq. (\ref{5}) and using Eq. (\ref{11}) we obtain:
\begin{eqnarray}
    \omega_{\Lambda}&=&-\frac{M_p^2/3}{3\alpha M_p^2+\gamma_1R \log\left(M_p^2/R\right)+\gamma_2R} +\nonumber\\
    && \frac{\left(1+ \Omega _k  \right)}{3\Omega _{\Lambda}}.\label{22}
\end{eqnarray}
We now want to derive the expression for the evolution of energy density parameter $\Omega_{\Lambda}$.\\
From Eq. (\ref{13}), it is possible to obtain the following expression for the EoS parameter  $\omega_{\Lambda}$:
\begin{eqnarray}
    \omega_{\Lambda}= -1-\frac{\dot{\rho}_{\Lambda}}{3H\rho_{\Lambda}}-\frac{Q}{3H\rho_{\Lambda}}.\label{23}
\end{eqnarray}
Using the expression of $Q$ given in Eq. (\ref{15}), the derivative of the DE density $\rho_{\Lambda}$ can be written as:
\begin{eqnarray}
    \dot{\rho}_{\Lambda}=3H\left[ -\rho_{\Lambda}-\left(\rho_m + \rho_{\Lambda}\right)\left(b^2+\frac{1}{3} \right) + \frac{RM_p^2}{3} \right].\label{24}
\end{eqnarray}
Dividing by the critical density $\rho_c=3H^2M_p^2$, Eq. (\ref{24}) can be written as:
\begin{eqnarray}
    \frac{\dot{\rho}_{\Lambda}}{\rho_c}&&=\dot{\Omega_{\Lambda}}+2\Omega_{\Lambda}\frac{\dot{H}}{H}\nonumber\\
    &&=3H\left[ -\Omega_{\Lambda}-\left(1 + \Omega_k\right)\left(b^2+\frac{1}{3} \right)
    + \frac{R}{9H^2} \right].  \label{25}
\end{eqnarray}
From Eq. (\ref{4}), we can see that the term $\frac{R}{9H^2}$ is equivalent to:
\begin{eqnarray}
    \frac{R}{9H^2}=\frac{2}{3}\left( \frac{\dot{H}}{H^2} +2 + \Omega_k \right).\label{26}
\end{eqnarray}
From Eq. (\ref{25}) and substituting Eq. (\ref{26}), it is possible to obtain the derivative of $\Omega_{\Lambda}$ with respect to the cosmic time $t$:
\begin{eqnarray}
\dot{\Omega}_{\Lambda}&=&2\frac{\dot{H}}{H}\left(1-\Omega_{\Lambda}
\right)+\nonumber\\
&&3H\left[-\Omega_{\Lambda}-  \left(1 +
\Omega_k\right)\left(b^2-\frac{1}{3} \right) + \frac{2}{3}
\right].\label{27}
\end{eqnarray}
Since   $\Omega_{\Lambda}'=\frac{d\Omega_{\Lambda}}{dx}= \frac{1}{H}\dot{\Omega}_{\Lambda}$ (where $x=\ln a$), we derive:
\begin{eqnarray}
H   \Omega_{\Lambda}'&=&2H'\left(1-\Omega_{\Lambda}
\right)+\nonumber\\
&&3H\left[-\Omega_{\Lambda}-  \left(1 +
\Omega_k\right)\left(b^2- \frac{1}{3} \right) + \frac{2}{3}
\right],\label{28}
\end{eqnarray}
which yields to:
\begin{eqnarray}
    \Omega_{\Lambda}'&=&\frac{2}{H}\left(1-\Omega_{\Lambda}  \right)+\nonumber\\
    &&3\left[-\Omega_{\Lambda}-  \left(1 + \Omega_k\right)\left(b^2
    -\frac{1}{3} \right) + \frac{2}{3} \right].\label{29}
\end{eqnarray}
In the above Equation we used the fact that:
\begin{eqnarray}
    H'=\frac{a'}{a}=1.\label{30}
\end{eqnarray}
For completeness, we also derive the deceleration parameter $q$:
\begin{eqnarray}
    q=-\frac{\ddot{a}a}{\dot{a}^2}=-1-\frac{\dot{H}}{H^2}.\label{31}
\end{eqnarray}
$q$, combined with the Hubble parameter $H$ and the dimensionless density parameters, form a set of very useful parameters for the description of the
astrophysical observations. Taking the derivative respect to the cosmic time $t$ of the Friedmann Eq. (\ref{7}) and using Eqs. (\ref{11}), (\ref{13})
and (\ref{14}), it is possible to write Eq. (\ref{31}) as:
\begin{eqnarray}
    q=\frac{1}{2}\left[1 + \Omega_k + 3\Omega_{\Lambda} \omega_{\Lambda}  \right].\label{32}
\end{eqnarray}
Substituting the expression of the EoS parameter $\omega_{\Lambda}$ given in Eq. (\ref{22}), we obtain that:
\begin{eqnarray}
    q=1-\frac{1}{2}\frac{M_p^2\Omega _{\Lambda}}{3\alpha M_p^2+\gamma_1R \log\left(M_p^2/R\right)+\gamma_2R}+ \Omega _k. \label{33}
\end{eqnarray}
We can derive the important quantities of the R-ECHDE model in the limiting case, for a flat dark dominated universe in HDE model, i.e.
when $\gamma_1=\gamma_2=0$, $\Omega_{\Lambda}=1$ and $\Omega_k=0$.\\
The energy density given in Eq. (\ref{5}) reduces to:
\begin{eqnarray}
\rho_{\Lambda}=3\alpha M_{p}^2 R.\label{34}
\end{eqnarray}
From FRW Eq. (\ref{7}) we find:
\begin{eqnarray}
H=\frac{6\alpha}{12\alpha-1}\left(\frac{1}{t}\right),\label{35}
\end{eqnarray}
\begin{eqnarray}
R=\frac{36\alpha}{(12\alpha-1)^2}\left(\frac{1}{t^2}\right).\label{36}
\end{eqnarray}
At last, the EoS parameter $\omega_{\Lambda}$ and deceleration parameter $q$ reduce to:
\begin{eqnarray}
\omega_{\Lambda}=\frac{1}{3}-\frac{1}{9\alpha},\label{LEoS}
\end{eqnarray}
\begin{eqnarray}
q=1-\frac{1}{6\alpha}.\label{Lq}
\end{eqnarray}
From Eq. (\ref{LEoS}), we see that in the limiting case, the EoS parameter of DE becomes a constant value in which for $\alpha<1/12$, $\omega_{\Lambda}<-1$,
where the phantom divide can be crossed. Since the Ricci scalar diverges at $\alpha=1/12$, this value of $\alpha$ can not be taken into account.
From Eq. (\ref{Lq}), the acceleration is started at $\alpha\leq 1/6$ where the quintessence regime is started ($\omega_{\Lambda} \leq -1/3$).\\
This case is very similar to power-law expansion of scale factor of
\citep{granda2008}, in which $a(t)=t^{6\alpha/(12\alpha-1)}$.

\section{CORRESPONDENCE BETWEEN R-ECHDE AND SCALAR FIELDS}
In this Section we establish a correspondence between the interacting Ricci scale model and the tachyon, K-essence, dilaton and quintessence
scalar field models. The importance of this correspondence is that the scalar field models are an effective description of an underlying theory
of DE. Therefore, it is worthwhile to reconstruct the potential and the dynamics of scalar fields according the evolutionary form of Ricci scalar model.
For this purpose, first we compare the energy density of Ricci scale model (i.e. Eq. (\ref{5})) with the energy density of corresponding scalar field model.
Then, we equate the equations of state of scalar field models with the EoS parameter of Ricci scalar model (i.e. Eq. (\ref{22})).

\subsection{\textbf{Interacting tachyon model}}
Recently, huge interest has been devoted to the study of the inflationary model with the help of the tachyon field, since it is believed the tachyon
can be assumed as a possible source of DE \citep{bagla, shao, calcagni, copeland-02-2005}. \\
The tachyon is an unstable field which can be used in string theory
through its role in the Dirac-Born-Infeld (DBI) action to describe
the D-bran action
\citep{sen-07-2002,sen-10-1999,bergshoeff,klusovn,kutasov}. Tachyon
might be responsible for cosmological inflation in the early
evolutionary stage of the universe, due to tachyon condensation near
the top of the effective scalar potential. A rolling tachyon has an
interesting EoS whose parameter smoothly interpolates between $-1$
and 0. This discovery motivated to take DE as a dynamical quantity,
i.e. a variable cosmological constant and model inflation using
tachyons. The effective Lagrangian for the tachyon field is given
by:
\begin{eqnarray}
L=-V(\phi)\sqrt{1-g^{\mu \nu}\partial _{\mu}\phi \partial_{\nu}\phi},\label{39}
\end{eqnarray}
where $V(\phi)$ represents the potential of tachyon and $g^{\mu \nu}$ is the metric tensor. The energy density $\rho_{\phi}$ and pressure $p_{\phi}$
for the tachyon field are given, respectively, by:
\begin{eqnarray}
    \rho_{\phi}&=&\frac{V(\phi)}{\sqrt{1-\dot{\phi}^2}},\label{40}\\
p_{\phi}&=& -V(\phi)\sqrt{1-\dot{\phi}^2}.\label{41}
\end{eqnarray}
Instead, the EoS parameter of tachyon scalar field is given by:
\begin{eqnarray}
w_{\phi}=\frac{p_{\phi}}{\rho_{\phi}}=\dot{\phi}^2-1.\label{42}
\end{eqnarray}
In order to have a real energy density for tachyon field, it is required that $-1 < \dot{\phi} < 1$. Consequently, from Eq. (\ref{42}),
the EoS parameter of tachyon is constrained to $-1 < \omega_{\phi} < 0$. Hence, the tachyon field can interpret the accelerated expansion of
the universe, but it can not enter the phantom regime, i.e. $\omega_{\Lambda}<-1$.\\
Comparing Eqs. (\ref{5}) and (\ref{40}), we obtain an expression for the potential $V\left( \phi \right)$ of the tachyon:
\begin{eqnarray}
    V(\phi)=\rho_{\Lambda} \sqrt{1-\dot{\phi}^2}.\label{43}
\end{eqnarray}
Instead, equating Eqs. (\ref{22}) and (\ref{42}), we obtain an expression for the kinetic energy term $\dot{\phi}^2$:
\begin{eqnarray}
&&\dot{\phi}^2= 1 +\omega_{\Lambda}=\label{44}\\
&&1-\frac{M_p^2/3}{3\alpha M_p^2+\gamma_1R
\log\left(M_p^2/R\right)+\gamma_2R} + \frac{\left(1+ \Omega _k
\right)}{3\Omega _{\Lambda}}.\nonumber
\end{eqnarray}
Using Eqs. (\ref{43}) and (\ref{44}), it is possible to write the potential of the tachyon as:
\begin{eqnarray}
    &&V\left( \phi  \right) = \rho _{\Lambda} \sqrt{-\omega_{\Lambda}} = \label{45} \\
    && \rho _{\Lambda} \sqrt{\frac{M_p^2/3}{3\alpha M_p^2+\gamma_1R \log\left(M_p^2/R\right)+\gamma_2R} -
    \frac{\left(1+ \Omega _k  \right)}{3\Omega _{\Lambda}}}.\nonumber
\end{eqnarray}
We can derive from Eqs. (\ref{44}) and (\ref{45}) that the kinetic energy $\dot{\phi}^2$ and the potential $V\left( \phi  \right)$
may exist if it is satisfied the condition:
\begin{eqnarray}
    -1\leq \omega_{\Lambda} \leq 0,\label{46}
\end{eqnarray}
which implies that the phantom divide can not be crossed in a universe with an accelerated expansion.\\
Using $\dot{\phi}=\phi'H$ and Eq. (\ref{44}), we get:
\begin{eqnarray}
 &&\phi'= \frac{1}{H}\times\label{47}\\
 &&  \sqrt{1-\frac{M_p^2/3}{3\alpha M_p^2+\gamma_1R \log\left(M_p^2/R\right)+\gamma_2R}
 + \frac{\left(1+ \Omega _k  \right)}{3\Omega _{\Lambda}}}.\nonumber
\end{eqnarray}
Then, from Eq. (\ref{47}), it is possible to obtain the evolutionary form of the tachyon scalar field as:
\begin{eqnarray}
    &&\phi\left(a\right) - \phi\left(a_0\right)=\int_{a_0}^a \frac{da}{aH}\times\label{48}\\
    &&\sqrt{1-\frac{M_p^2/3}{3\alpha M_p^2+\gamma_1R \log\left(M_p^2/R\right)+\gamma_2R} + \frac{\left(1+ \Omega _k  \right)}{3\Omega _{\Lambda}}},\nonumber
\end{eqnarray}
where $a_0$ is the present value of the scale factor. Here, we have established an interacting
entropy-corrected holographic tachyon DE model and reconstructed the potential
and the dynamics of the tachyon field.\\
In the limiting case for flat dark dominated universe for $\gamma_1=\gamma_2=0$, $\Omega_{\Lambda}=1$ and $\Omega_k$=0, the scalar field and
potential of the tachyon are, respectively:
\begin{eqnarray}
\phi(t)=\sqrt{\frac{12\alpha -1}{9\alpha}}t,\label{49}
\end{eqnarray}
\begin{eqnarray}
V(\phi)=\frac{4M_p^2}{(12\alpha-1)}\sqrt{\alpha(1-3\alpha)}\frac{1}{\phi^2},\label{50}
\end{eqnarray}
which are a result of the power-law expansion. In this correspondence, the scalar field exist provided that $\alpha >1/12$, which shows that the
phantom divide can not be achieved.

\subsection{\textbf{Interacting K-essence model}}
A model in which the kinetic term of the scalar field appears in the
Lagrangian in a non-canonical way is called K-essence model. The
idea of the K-essence scalar field was motivated from the
Born-Infeld action of string theory and it is used to explain the
late time acceleration of the universe \citep{sen-2002,lambert}. The
general scalar field action for the K-essence model as a function of
$\phi$  and $\chi=\dot{\phi}/2$ is given by
\citep{chiba-07-2000,armendariz-11-2000,armendariz-05-2001}:
\begin{eqnarray}
S=\int d^4x \sqrt{-g}\,p\left(\phi, \chi \right),\label{51}
\end{eqnarray}
where the Lagrangian density $p\left(\phi, \chi \right)$ corresponds to a pressure density. According to this Lagrangian, the pressure $p\left(\phi, \chi \right)$ and the energy density of the field $\phi$ can be written, respectively, as:
\begin{eqnarray}
    p\left(\phi, \chi \right)&=&f\left(\phi\right)\left( -\chi+\chi ^2   \right), \label{52}\\
        \rho\left(\phi, \chi \right)&=&f\left(\phi\right)\left(-\chi+3\chi ^2\right).\label{53}
\end{eqnarray}
Hence, the EoS parameter of K-essence scalar field is given by:
\begin{eqnarray}
    \omega _K= \frac{p\left(\phi, \chi \right)}{\rho\left(\phi, \chi \right)}=\frac{\chi-1}{3\chi -1}.\label{54}
\end{eqnarray}
From Eq. (\ref{54}), one can see the phantom behavior of K-essence scalar field ($w_K < -1$) when the parameter $\chi$ lies in the interval $1/3 < \chi < 1/2$.\\
In order to consider the K-essence field as a description of the interacting R-ECHDE density, we establish the correspondence between the K-essence EoS parameter,
$\omega_K$, and the interacting R-ECHDE EoS parameter, $\omega_{\Lambda}$.\\
The expression for $\chi$ can be found equating  Eqs. (\ref{22}) and (\ref{54}), obtaining:
\begin{eqnarray}
   && \chi = \frac{w_{\Lambda}-1}{3w_{\Lambda}-1}=\nonumber\\
   &&\frac{-1-\frac{M_p^2/3}{3\alpha M_p^2+\gamma_1R \log\left(M_p^2/R\right)+\gamma_2R} + \frac{\left(1+ \Omega _k  \right)}{3\Omega _{\Lambda}}}{-1- \frac{M_p^2}{3\alpha M_p^2+\gamma_1R \log\left(M_p^2/R\right)+\gamma_2R} + \frac{\left(1+ \Omega _k  \right)}{\Omega _{\Lambda}}}.\label{55}
\end{eqnarray}
Moreover, equating Eqs. (\ref{5}) and (\ref{53}), we obtain:
\begin{eqnarray}
    f\left(\phi \right)=\frac{\rho_{\Lambda}}{\chi(3\chi-1)}.\label{56}
\end{eqnarray}
Using $\dot{\phi}^2=2\chi$ and $\dot{\phi}=\phi'H$, we derive:
\begin{eqnarray}
 \phi'= \frac{\sqrt{2}}{H}\sqrt{\frac{-1-\frac{M_p^2/3}{3\alpha M_p^2+\gamma_1R \log\left(M_p^2/R\right)+\gamma_2R} + \frac{\left(1+ \Omega _k  \right)}
 {3\Omega _{\Lambda}}}{-1- \frac{M_p^2}{3\alpha M_p^2+\gamma_1R \log\left(M_p^2/R\right)+\gamma_2R} + \frac{\left(1+ \Omega _k  \right)}{\Omega _{\Lambda}} }}.\label{57}
\end{eqnarray}
Integrating Eq. (\ref{57}), we find the evolutionary form of the K-essence scalar field:
\begin{eqnarray}
    &&\phi\left(a\right) -    \phi\left(a_0\right) = \sqrt{2} \int_{a_0}^a \frac{da}{aH}\times\nonumber\\
    &&\sqrt{\frac{-1-\frac{M_p^2/3}{3\alpha M_p^2+\gamma_1R \log\left(M_p^2/R\right)+\gamma_2R} + \frac{\left(1+ \Omega _k  \right)}{3\Omega _{\Lambda}}}{-1- \frac{M_p^2}{3\alpha M_p^2+\gamma_1R \log\left(M_p^2/R\right)+\gamma_2R} + \frac{\left(1+ \Omega _k  \right)}{\Omega _{\Lambda}} }} ,\label{58}
\end{eqnarray}
where $a_0$ is the present value of the scale factor.\\
In the limiting case of $\gamma_1=\gamma_2=0$, $\Omega_{\Lambda}=1$ and $\Omega_k$=0, in a flat dark dominated universe, the scalar field and potential
of K-essence field reduce to:
\begin{eqnarray}
\phi(t)=\sqrt{\frac{2(1+6\alpha)}{3}}t,\label{59}
\end{eqnarray}
\begin{eqnarray}
f(\phi)=\frac{36\alpha
M_p^2}{(12\alpha-1)^2}\frac{1}{\phi^2},\label{60}
\end{eqnarray}
which are a result of power-law expansion. Moreover, we see that our universe may behave in all accelerated regimes (phantom and quintessence),
since all values of $\alpha$ are permitted.

\subsection{\textbf{Interacting dilaton model}}
A dilaton scalar field, originated from the lower-energy limit of string theory \citep{piazza}, can also be assumed as a source of DE.\\
The process of compactification of the string theory from higher to
four dimensions introduces the scalar dilaton field which is coupled
to curvature invariants. The coefficient of the kinematic term of
the dilaton can be negative in the Einstein frame, which means that
the dilaton behaves as a phantom-like scalar field. The pressure
(Lagrangian) density and the energy density of the dilaton DE model
are given, respectively, by
\citep{gasperini,arkani,elizalde-02-2008}:
\begin{eqnarray}
p_D&=&-\chi +ce ^{\lambda \phi}\chi^2, \label{61}  \\
\rho_D&=&-\chi +3ce ^{\lambda \phi}\chi^2,\label{62}
\end{eqnarray}
where $c$ and $\lambda$ are positive constants and $2\chi=\dot{\phi}^2$. The negative coefficient of the kinematic term of the dilaton field in Einstein
frame makes a phantom-like behavior for dilaton field. The EoS parameter for the dilaton scalar field is given by:
\begin{eqnarray}
    \omega _D= \frac{p_D}{\rho_D}=\frac{-1 +c e ^{\lambda \phi}\chi}{-1 +3c e ^{\lambda \phi}\chi}.\label{63}
\end{eqnarray}
In order to consider the dilaton field as a description of the interacting R-ECHDE density, we establish the correspondence between the
 dilaton EoS parameter, $w_D$, and the EoS parameter $\omega_{\Lambda}$ of the R-ECHDE model. By equating Eqs. (\ref{22}) and (\ref{63}), it is possible to find:
\begin{eqnarray}
    &&ce ^{\lambda \phi}\chi =\frac{\omega_{\Lambda}-1}{3\omega_{\Lambda}-1}=\nonumber\\
    &&\frac{-1-\frac{M_p^2/3}{3\alpha M_p^2+\gamma_1R \log\left(M_p^2/R\right)
    +\gamma_2R} + \frac{\left(1+ \Omega _k  \right)}{3\Omega _{\Lambda}}}{-1- \frac{M_p^2}{3\alpha M_p^2
    +\gamma_1R \log\left(M_p^2/R\right)+\gamma_2R} + \frac{\left(1+ \Omega _k  \right)}{\Omega
    _{\Lambda}}}. \label{64}
    \end{eqnarray}
Using $\dot{\phi}^2=2\chi$, we can rewrite Eq. (\ref{64}) as:
\begin{eqnarray}
    &&e^{\lambda \phi/2} \dot{\phi}=\sqrt{\frac{2}{c}}\times\nonumber\\
    &&\sqrt{ \frac{-1-\frac{M_p^2/3}{3\alpha M_p^2+\gamma_1R \log\left(M_p^2/R\right)+\gamma_2R}
     + \frac{\left(1+ \Omega _k  \right)}{3\Omega _{\Lambda}}}   {-1- \frac{M_p^2}{3\alpha M_p^2+\gamma_1R \log\left(M_p^2/R\right)
     +\gamma_2R} + \frac{\left(1+ \Omega _k  \right)}{\Omega _{\Lambda}}}}.\label{65}
\end{eqnarray}
Integrating Eq. (\ref{65}) with respect to $a$, we obtain:
\begin{eqnarray}
    &&e^{\frac{\lambda \phi\left(a\right)}{2}} = e^{\frac{\lambda \phi\left(a_0\right)}{2}}+\frac{\lambda}{\sqrt{2c}}\int_{a_0}^a\frac{da}{aH}\times \nonumber \\
    &&\sqrt{\frac{-1-\frac{M_p^2/3}{3\alpha M_p^2+\gamma_1R \log\left(M_p^2/R\right)+\gamma_2R} + \frac{\left(1+ \Omega _k  \right)}{3\Omega _{\Lambda}}}   {-1- \frac{M_p^2}{3\alpha M_p^2+\gamma_1R \log\left(M_p^2/R\right)+\gamma_2R} + \frac{\left(1+ \Omega _k  \right)}{\Omega _{\Lambda}}}}.\label{66}
\end{eqnarray}
The evolutionary form of the dilaton scalar field is written as:
\begin{eqnarray}
    &&\phi\left(a\right)=  \frac{2}{\lambda}\ln {\left[e^{\frac{\lambda \phi\left(a_0\right)}{2}}+\right]}+ \frac{\lambda}{\sqrt{2c}}  \int_{a_0}^a \frac{da}{aH}\times  \nonumber\\
   &&\sqrt{ \frac{-1-\frac{M_p^2/3}{3\alpha M_p^2+\gamma_1R \log\left(M_p^2/R\right)
   +\gamma_2R} + \frac{\left(1+ \Omega _k  \right)}{3\Omega _{\Lambda}}}   {-1- \frac{M_p^2}{3\alpha M_p^2+\gamma_1R \log\left(M_p^2/R\right)+\gamma_2R}
    + \frac{\left(1+ \Omega _k  \right)}{\Omega _{\Lambda}}}}.\label{67}
\end{eqnarray}
In the limiting case of $\gamma_1=\gamma_2=0$, $\Omega_{\Lambda}=1$ and $\Omega_k$=0, in a flat dark dominated universe, the scalar field of dilaton field reduces to the following form:
\begin{eqnarray}
\phi(t)=\frac{2}{\lambda}\ln{\left[\lambda t\sqrt{\frac{1+6\alpha}{6c}}\right]}.\label{68}
\end{eqnarray}
We see that all values of $\alpha$ are permitted and, therefore, by this correspondence, the universe may behave in phantom and quintessence regime.

\subsection{\textbf{Quintessence}}
Quintessence is described by an ordinary time-dependent and
homogeneous scalar field $\phi$ which is minimally coupled to
gravity, but with a particular potential $V\left(\phi\right)$ that
leads to the accelerating universe. The action for quintessence is
given by \citep{copeland-2006}:
\begin{eqnarray}
    S=\int d^4x \sqrt{-g}\,\left[-\frac{1}{2}g^{\mu \nu} \partial _{\mu} \phi   \partial _{\nu} \phi - V\left( \phi \right)  \right].\label{69}
\end{eqnarray}
The energy momentum tensor $T_{\mu \nu}$ of the field is derived by varying the action given in Eq. (\ref{69}) with respect to the metric tensor $g^{\mu \nu}$:
\begin{eqnarray}
T_{\mu \nu}=\frac{2}{\sqrt{-g}} \frac{\delta S}{\delta g^{\mu \nu}},\label{70}
\end{eqnarray}
which yields to:
\begin{eqnarray}
    T_{\mu \nu}=\partial _{\mu} \phi   \partial _{\nu} \phi - g_{\mu \nu}\left[\frac{1}{2}g^{\alpha \beta} \partial _{\alpha} \phi   \partial _{\beta} \phi + V\left( \phi \right)  \right].\label{71}
\end{eqnarray}
The energy density $\rho_Q$ and pressure $p_Q$ of the quintessence scalar field $\phi$ are given, respectively, by:
\begin{eqnarray}
    \rho_Q&=&-T_0^0=\frac{1}{2}\dot{\phi}^2+V\left(\phi\right),\label{72}\\
    p_Q&=&T_i^i=\frac{1}{2}\dot{\phi}^2-V\left(\phi\right).\label{73}
\end{eqnarray}
The EoS parameter for the quintessence scalar field is given by:
\begin{eqnarray}    \omega_Q=\frac{p_Q}{\rho_Q}=\frac{\dot{\phi}^2-2V\left(\phi\right)}{\dot{\phi}^2+2V\left(\phi\right)}.\label{74}
\end{eqnarray}
We find from Eq. (\ref{74}) that, when $\omega_Q < -1/3$, the universe accelerates for $\dot{\phi}^2<V\left(\phi\right)$.\\
Here we establish the correspondence between the interacting scenario and the quintessence DE model: equating Eq. (\ref{74}) with the EoS parameter given in
 Eq. (\ref{22}), i.e.  $\omega_Q=\omega_{\Lambda}$, and equating Eqs. (\ref{72}) and(\ref{5}), i.e.  $\rho_Q=\rho_{\Lambda}$, we obtain:
\begin{eqnarray}
    \dot{\phi}^2&=&\left(1+\omega_{\Lambda} \right)\rho_{\Lambda}, \label{75}\\
    V\left( \phi \right) &=& \frac{1}{2}\left(1-\omega_{\Lambda} \right)\rho_{\Lambda}.\label{76}
\end{eqnarray}
Substituting Eq. (\ref{23}) into Eqs. (\ref{75}) and (\ref{76}), the kinetic energy term $\dot{\phi}^2$ and the quintessence potential energy
 $V\left( \phi \right)$ can be easily found as follow:
\begin{eqnarray}
    \dot{\phi}^2&=&\rho_{\Lambda}\Big{(}1-\frac{M_p^2/3}{3\alpha M_p^2+\gamma_1R \log\left(M_p^2/R\right)+\gamma_2R} +\nonumber\\
    && \frac{\left(1+ \Omega _k \right)}{3\Omega _{\Lambda}}\Big{)},\label{77}\\
    V\left( \phi \right) &=&\frac{\rho_{\Lambda}}{2}  \Big{(} 1+\frac{M_p^2/3}{3\alpha M_p^2+\gamma_1R
    \log\left(M_p^2/R\right)+\gamma_2R}-\nonumber\\
    && - \frac{\left(1+ \Omega _k  \right)}{3\Omega _{\Lambda}}  \Big{)}.\label{78}
\end{eqnarray}
From Eq. (\ref{77}), using $\dot{\phi}=\phi' H$, it is possible to obtain the evolutionary form of the quintessence scalar field as:
\begin{eqnarray}
&&\phi\left(a\right) - \phi \left(a_0\right)= \int_{a_0}^{a}\frac{da}{a}\Big{\{}\sqrt{3M_p^2\Omega_{\Lambda}}\times \nonumber \\
&&\Big{(}1-\frac{M_p^2/3}{3\alpha M_p^2+\gamma_1R
\log{(M_p^2/R)}+\gamma_2R} +\nonumber\\
&&\frac{\left(1+ \Omega _k \right)}{3\Omega
_{\Lambda}}\Big{)}^{1/2}\Big{\}},\label{79}
\end{eqnarray}
where $a_0$ is the present value of the scale factor. In the limiting case of $\gamma_1=\gamma_2=0$, $\Omega_{\Lambda}=1$ and $\Omega_k$=0,
in a flat dark dominated universe, the scalar field and potential of quintessence reduce to:
\begin{eqnarray}
\phi(t)&=&\frac{6\alpha M_p}{\sqrt{3\alpha(12\alpha-1)}}\ln{(t)},\label{80}\\
V(\phi)&=&\frac{6\alpha(6\alpha+1)}{(12\alpha-1)^2}M_p^2\exp{\left[\frac{-\sqrt{3\alpha(12\alpha-1)}}{3\alpha
M_p}\phi\right]}.\label{81}
\end{eqnarray}
The potential exists for all values of $\alpha >1/12$ (quintessence
regime). The potential has also been obtained by power-law expansion
of scale factor.

\section{Conclusions}
In this paper, we considered the entropy-corrected version of the
HDE model which is in interaction with DM in the non-flat FRW
universe (and with IR cut-off equivalent to the Ricci scalar R). The
HDE model is an attempt to probe the nature of DE within the
framework of quantum gravity. We considered the logarithmic
correction term to the energy density of HDE model. The addition of
correction terms to the energy density of HDE is motivated from the
Loop Quantum Gravity (LQG), which is one of the most promising
theories of quantum gravity. Using the expression of this modified
energy density, we obtained the EoS parameter, deceleration
parameter and evolution of energy density parameter for the
interacting R-ECHDE model. We found that for the appropriate model
parameters (even in limiting case,$\gamma_1=\gamma_2=\Omega_k=0$,
$\Omega_{\Lambda}=1$), the phantom divide may be crossed,
$\omega_{\Lambda}<-1$, and the present acceleration expansion
($q<0$) is achieved where the  quintessence regime is started.
Moreover, we established a correspondence between the interacting
R-ECHDE model and the tachyon, K-essence, dilaton and quintessence
scalar field models in the hypothesis of non-flat
FRW universe.\\
These correspondences are important to understand how various candidates of DE are mutually related to each other. The limiting case of flat
dark dominated universe without entropy correction were studied in each scalar field and we see that the EoS parameter is constant in this case
and we calculate the scalar field and its potential which can be obtained by idea of power-law expansion of scalar field.\\
In order to make a comparison between our model and another works in
LECHDE-scalar field model, we concentrate our attention in a recent
article \citep{amani}. The authors considered a scalar-tensor
cosmological model with the non-minimal kinetic coupling terms and
discussed its cosmological implications with respect to the entropy
corrected holographic dark energy. Our results differ from their
results in that their analysis involves two coupling parameters and
a cosmological event horizon while ours deal with a Ricci scale and
no couplings. Such scalar field models have interesting property of
explaining the phantom crossing while the reconstructed scalar
potential has interesting physical implications in cosmology.
\\

\acknowledgments {We are grateful to the referee for valuable
comments and suggestions, which have allowed us to improve this
paper significantly. Also M. Jamil would like to thank the warm
hospitality of the Abdus Salam ICTP, Trieste where this work was
initiated.}

\end{document}